\documentstyle[12pt,epsf,epsfig]{article}
\def\beq{\begin{equation}}
\def\eeq{\end{equation}}
\def\beqa{\begin{eqnarray}}
\def\eeqa{\end{eqnarray}}
\def\d{{\rm d}}

\begin{document}

\begin{flushright}
YITP--SB--01--65
\end{flushright}

\vspace*{0.1in}

\centerline{\large \bf
QCD and Rescattering in Nuclear Targets}
\vspace*{.4in}

\centerline{Jianwei Qiu$^a$ and George Sterman$^b$}
\vspace*{.2in}

\centerline{\small \it $^a$ Department of Physics and Astronomy,
Iowa State University}

\centerline{\small \it Ames, Iowa 50011}

\centerline{\small \it $^b$C.N. Yang Institute for
Theoretical Physics, State University of
New York}

\centerline {\small \it Stony Brook, NY 11794-3840}
\vspace*{.3in}

\begin{abstract}
We review the extension of the factorization formalism
of perturbative QCD to soft rescattering associated with hard
processes in nuclei.
\end{abstract}

\vspace*{.1in}

\section{Introduction}

\begin{figure}[ht]
\centerline{\epsffile{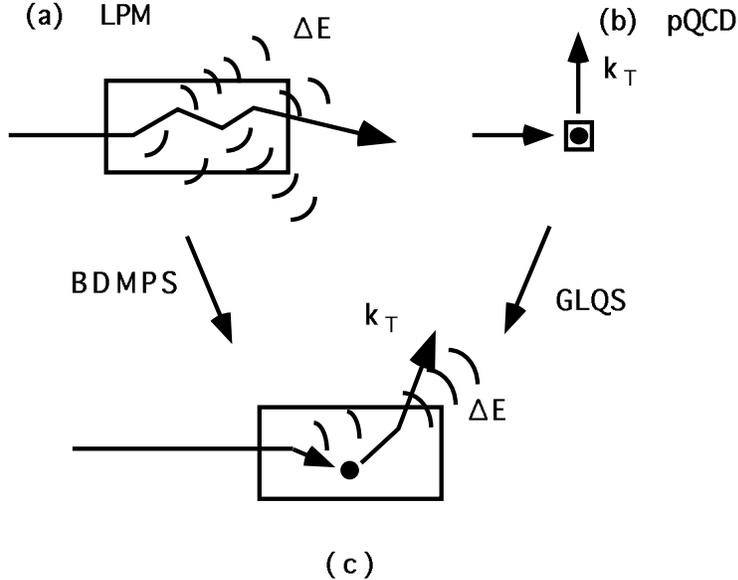}}
\caption{Alternate approaches to hard scattering in nuclei. (a)
Landau-Pomeranchuk-Migdal analysis treats energy loss due to many
soft scatterings. (b) Perturbative QCD analysis treats momentum
transfer due to hard scattering. (c)  For scattering in nuclei,
both must be combined.}
\label{fig1}
\end{figure}

This paper reviews a
perturbative QCD (pQCD) treatment of the
hard scattering of hadrons and
leptons in nuclei, based on factorization,
It describes in part work in collaboration with
Ma Luo \cite{LQS1,LQS,LQS3} and  Xiaofeng Guo
\cite{GQ,Guo:1998rd,Guo:2000eu}.
At the outset, it may be useful to clarify the relation of this work
to the works of Baier {\it et al.} (BDMPS) \cite{BDMPS} and
Zakharov \cite{Zakharov}.
For more recent progress in relating these two
approaches, see \cite{GW-loss}.
We have tried to illustrate this relation schematically
in Fig.~\ref{fig1}.  The BDMPS analysis begins (Fig.~\ref{fig1}a) with
the classic treatment of radiation induced when a charged particle
passes through a large target, due originally
to Landau, Pomeranchuk and Migdal (LPM).  This analysis does not require
the presence of a hard scattering, but describes the coherent results
of  many soft scatterings.  Its primary subject has traditionally been
induced energy loss.  Our analysis (GLQS) begins with the perturbative
QCD treatment of hard-scattering in a relatively small target
(Fig.~\ref{fig1}b), in which the primary subject of interest is
momentum transfer.  A complete analysis (Fig.~\ref{fig1}c) of hard
scattering in a large target involves both energy loss and the
transverse momenta due to initial- and final-state soft scatterings.
Our work is a step in this direction, attempting to stay as close as
possible to the pQCD formalism, in which we may readily quantify
corrections.  To be specific, we consider only a single soft initial-
or final-state interaction in addition to the hard scattering.
Our central observation is that for suitably-defined jet  and related
inclusive cross sections this is the first order in an
expansion in the quantity
\beq
{A^{1/3}\times\lambda^2\over Q^2}\, ,
\label{param}
\eeq
where $\lambda$ represents a nonperturbative scale,
which we shall identify  with a higher-twist
parton distribution below.
That additional scatterings are suppressed by factors of  $1/Q^2$ is
perhaps surprising.  Let us review why this is the case, at least for
certain cross sections.

\begin{figure}
\begin{minipage}[t]{2.6in}
\begin{center}
\epsfig{figure=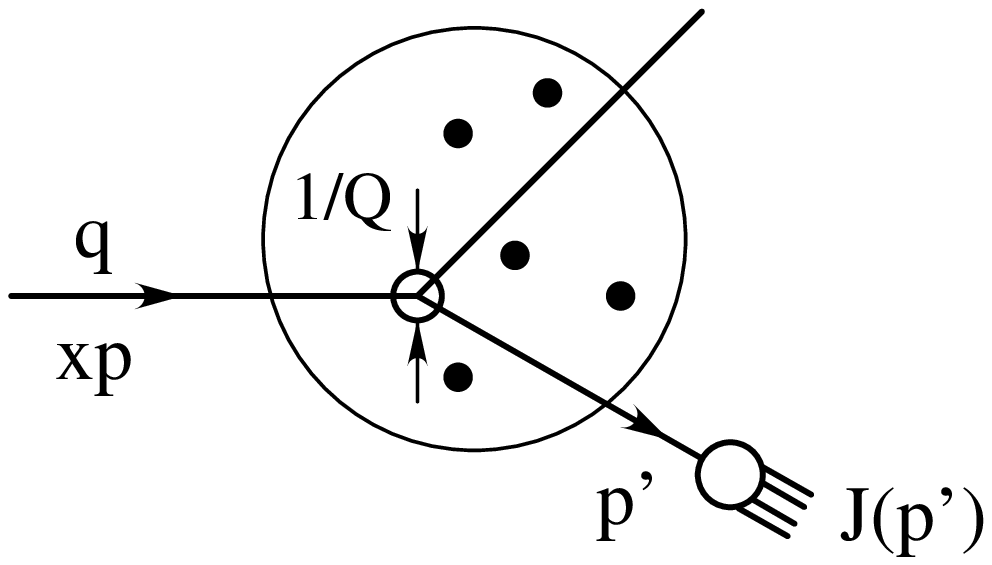,width=2.2in}
\end{center}
\caption{Sketch for the scattering of a quark of momentum $xp$ in a
large nucleus.}
\label{fig2}
\end{minipage}
\hfill
\begin{minipage}[t]{3.5in}
\begin{center}
\epsfig{figure=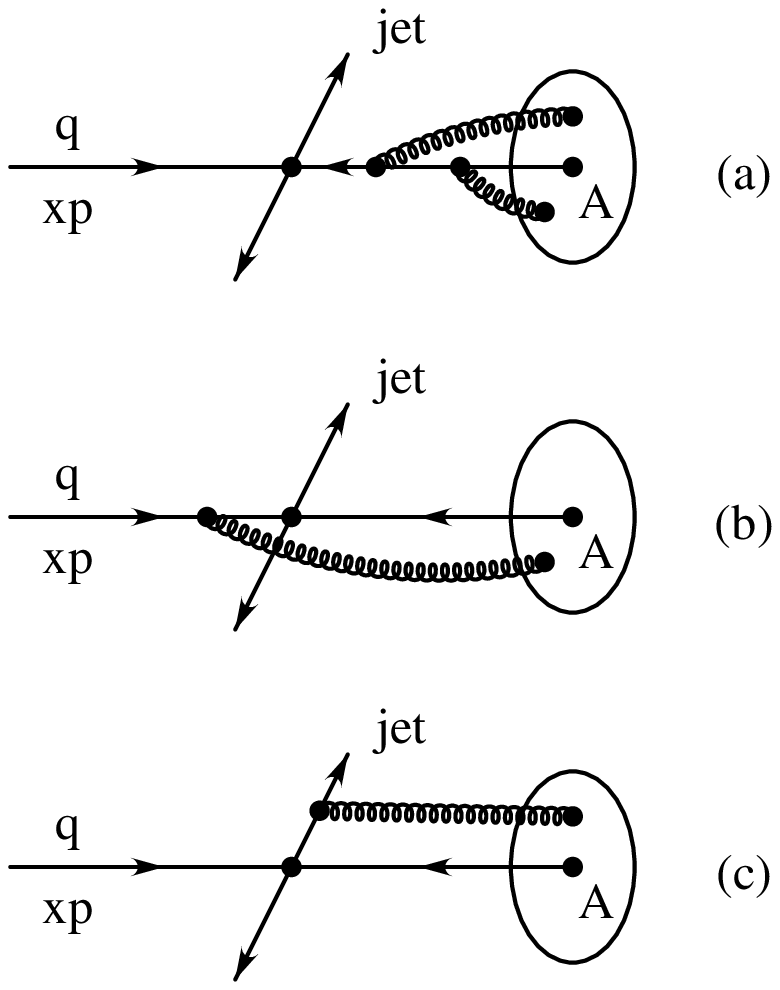,width=2.4in}
\end{center}
\caption{Classification of parton multiple scattering in nuclear
medium: (a) interactions internal to the nucleus, (b) initial-state
interactions, and (c) final-state interactions.}
\label{fig3}
\end{minipage}
\end{figure}

The basic analysis of hard-scattering in nuclear matter  (cold or hot)
\cite{tdilat} is quite simple.  To be specific, consider the
scattering of a quark, as shown in Fig.~\ref{fig2}.
A hard-scattering with momentum transfer $Q$
can resolve states whose lifetimes are as short as $1/Q$, for instance
quarks off-shell by order $Q$, but still less that $Q$.
The off-shellness of the scattered quark increases with the momentum
transfer simply because {\it the number of available states
increases with increasing momentum}.
Similarly, the scattered  quark, of momentum $p'$ is typically
off-shell by order $m_J\le Q$.
We may think of $m_J$ as  the invariant mass of
the jet into which quark fragments.
If we are to recognize the jet, we must have $m_J\ll E_J=p'_0$, with
$E_J$ being energy of the jet.  On the other hand, the counting of
available states ensures that $m_J\gg\Lambda_{\rm QCD}$.

Now the scattered quark
has a lifetime in its own rest frame $\Delta t^{(p')}
\sim {1\over m_J}$ with $m_J\ll E_J$.
In the target rest frame, however, this becomes,
for large enough $E_J/m_J$, $\Delta t^{\rm (target)}
\sim{1\over m_J}\left({E_J\over m_J}\right)>R_A$,
where $R_A$ is the (fixed) target size.  Thus, at high enough energy
the lifetime of the scattered quark will exceed the target size, even
though the quark itself is far off the mass shell, typically
by a scale that grows with the momentum transfer $Q$.

Further couplings of the off-shell quark are suppressed, first of all
by the strong coupling evaluated at scale $m_J$, and, more importantly,
by an overall factor of $1/m_J^2\sim 1/Q^2$, since the effective
size of the scattered quark decreases with momentum
transfer in this manner.

In summary, for inclusive processes such as jet production, high-$Q$
implies that process-dependent multiple
scattering is power-suppressed compared
to single scattering.
The most important point here is that the scattered particle
remains off-shell for its entire transit of the target.
Thus, its interactions with the target may be
treated by the formalism of  perturbative QCD,
which, however, must be extended to include corrections
that decrease with extra powers of momentum transfer.
Up to the first such ``higher-twist" contribution, a general
cross section has the representation \cite{QS}
\begin{equation}
\sigma(Q) = H^0\otimes f_2\otimes f_2 +
\left(\frac{1}{Q^2}\right) H^1\otimes f_2 \otimes f_{4}
+ O\left(\frac{1}{Q^4}\right)\, ,
\label{HTE}
\end{equation}
where $\otimes$ represents convolutions in fractional momenta
carried by partons, and $f_n$ represents a parton distribution of
twist $n$.  Target-size dependence due to multiple scattering can
only appear in the second term in this expansion.

\section{Parton-Nucleus Scattering in Perturbative QCD}

To distinguish parton-nucleus multiple scattering from partonic
dynamics internal to the nucleus, we classify the multiple
scattering internal to a nuclear target in the following three categories:
(a) initial-state interactions internal to the nucleus, (b)
initial-state parton-nucleus interactions (ISI), and (c) final-state
parton-nucleus interactions (FSI), as shown in Fig.~\ref{fig3}.
To a certain degree, this classification is ambiguous,
but it can be made well-defined if we are careful.


Initial-state interactions internal to the nucleus change the
twist-2 parton distributions of the nucleus, as shown in
Fig.~\ref{fig3}a.
As a result, the effective
parton distributions of a large nucleus are different from a
simple sum of individual nucleon's parton distributions.  This
is known as the ``EMC'' effect for the region where the
parton's momentum fraction $x$ is not too small.
Since only a single parton from the nucleus participates the hard
collision  to leading power,
the effect of the initial-state interactions
internal to the nucleus is (almost by definition) leading twist.
The $A$-dependence of the ``EMC'' effect provides a
relatively small nuclear size
dependence to the first term in the Eq.~(\ref{HTE}).

On the other hand, the initial-state and final-state
parton-nucleus interactions, as shown in Fig.~\ref{fig3}b and
\ref{fig3}c,
involve at least two physical partons from the nucleus at the hard
collisions.  Thus the ``Cronin effect", $A^\alpha$-dependence
with $\alpha>1$, due to multiple scattering, is higher-twist for
inclusive distributions.

\subsection{Factorization at Leading Powers}

Let us review some of the details of a factorized cross section like
the one in Eq.~(\ref{HTE}).
\begin{figure}
\begin{minipage}[t]{3in}
\begin{center}
\epsfig{figure=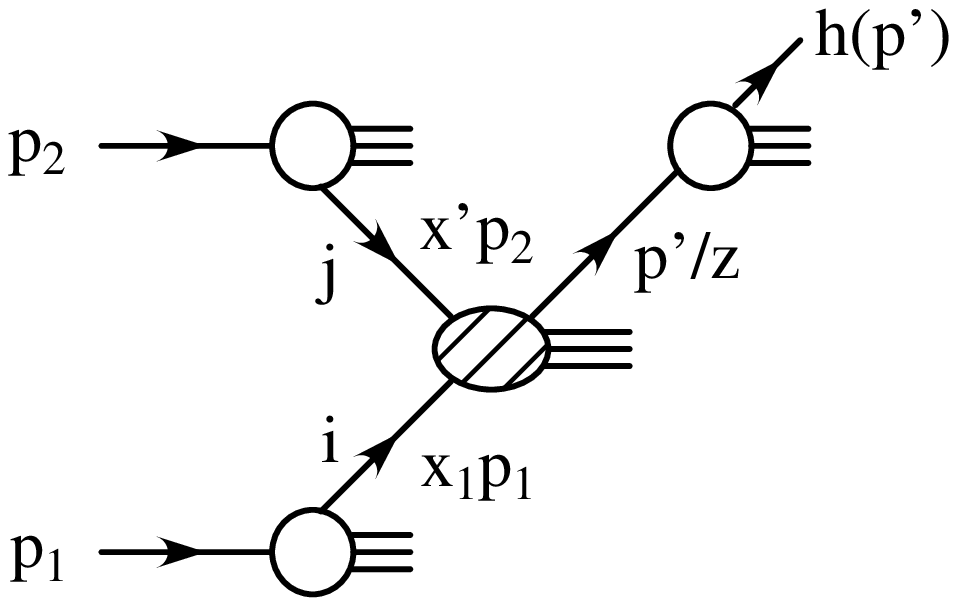,width=2.6in}
\end{center}
\caption{Scattering amplitude between two incoming hadrons with one
hard collision.}
\label{fig4}
\end{minipage}
\hfill
\begin{minipage}[t]{3in}
\begin{center}
\epsfig{figure=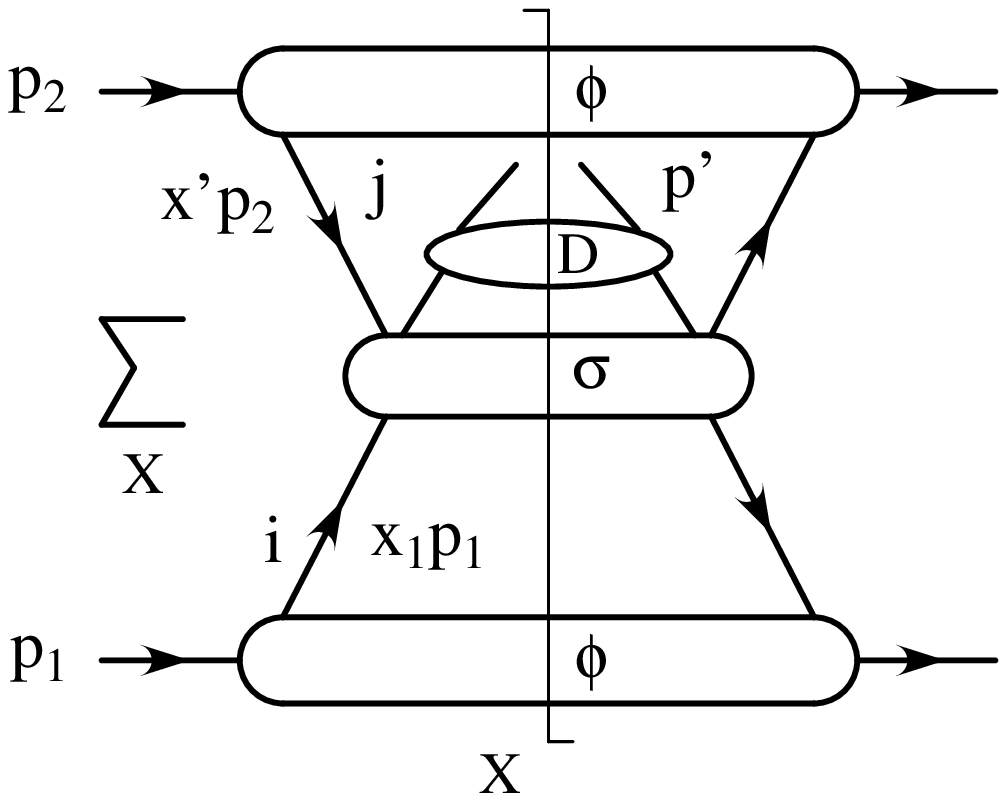,width=2.5in}
\end{center}
\caption{Perturbative QCD factorization at leading twist.}
\label{fig5}
\end{minipage}
\end{figure}

The leading power contributions to a general cross section involve
only one hard collision between two partons from two incoming hadrons
(or nuclei), as shown in Fig.~\ref{fig4}.  An energetic third parton
is produced in the collision, which fragments into either a jet or
a hadron of momentum $p'$.  After squaring the scattering amplitude in
Fig.~\ref{fig4}, and summing over all final-states, the cross section
can be factorized into a form like the first term in Eq.~(\ref{HTE}),
\beq
\omega{d\sigma_2\over d^3p'}
=
\sum_{ijk}\int\frac{dx'}{x'}f_{j/p_2}(x')\;
           \int\frac{dx_1}{x_1}f_{i/p_1}(x_1)\;
           \int\frac{dz}{z^2} D_{h/k}(z)\,
           \hat{\sigma}_{ij\rightarrow k}(x_1p_1,x'p_2,p'/z)\, ,
\label{twist2conv}
\eeq
where $\sum_{ijk}$ runs over all parton flavors and all scale
dependence is implicit.
The $D_{h/k}$ are fragmentation functions for a parton of type
$k$ to produce a hadron $h$.  For jet production, the
fragmentation from a parton to a jet, suitably defined, is calculable
in perturbation theory, and may be absorbed into the ``hard scattering
function'' $\hat{\sigma}$.  Then, the factorized single scattering
formula in Eq.~(\ref{twist2conv}) is reduced to the first term in
Eq.~(\ref{HTE}).
The $f_{a/p}$ are twist-2 distributions of parton type $a$  in
hadron $p$.  They have the interpretation of expectation
values in the hadronic state
of products of fields on the light cone, for
instance, for a quark distribution
\beq
f_{q/p}(x,Q)=
\int{dy^- \over 2\pi} {\rm e}^{ixp^+y^-}
\langle p|\bar{q}(0){\gamma^+\over 2}q(y^-)|p\rangle\, ,
\label{q-dis}
\eeq
where for simplicity we choose the $A^+=0$ gauge, assuming $\vec{p}$ is
in the plus direction.  Eq.~(\ref{twist2conv}) is illustrated
by Fig.~\ref{fig5}.
As shown, the convolution in Eq.~(\ref{twist2conv}) is in terms of the
momentum fractions $x_1$ and $x'$ carried by partons $i$ and  $j$,
from hadrons $p_1$ and $p_2$, respectively, into the hard scattering.

The factorized formula in Eq.~(\ref{twist2conv}) illustrates
the general leading power collinear factorization theorem
\cite{CSS-fac}.  It 
consistently separates perturbatively calculable short-distance
physics into $\hat{\sigma}$, and isolates long-distance effects
into universal nonperturbative matrix elements (or distributions),
such as $f_{a/p}$ or $D_{h/k}$, associated with each observed hadron.
Quantum interference between  long- and short-distance physics is
power-suppressed, by the large energy exchange of the collisions.
Predictions of pQCD follow when processes with different hard
scatterings but the same nonperturbative matrix elements are compared.

In the case of collisions on a nuclear target, the factorized single
scattering formula remains valid, except that the twist-2 parton
distribution $f_{a/p}$ is defined on a nuclear state, instead of a free
nucleon state.  For example, for a nucleus of momentum $P_A\equiv
A\, p_A$, the effective quark distribution is defined in the same
way as in Eq.~(\ref{q-dis}), with $|p\rangle$
a nuclear state, $|P_A\rangle$.  Such an effective
nuclear parton distribution includes the ``EMC'' effect, and is
still a twist-2 distribution function by the definition of its
operator.

Power-suppressed corrections to
Eq.~(\ref{twist2conv}) involve ratios of the
nonperturbative momentum scales in the hadron, $\lambda\sim
\Lambda_{\rm QCD}\sim 1/{\rm fm}$, over the energy exchange of hard
collisions, $Q$, as $(\lambda^2/Q^2)^n$.  These corrections can come
from several different sources, including the effect of partons'
non-collinear momentum components, and the effects of
interactions involving more than one partons from each
hadron.  In the case of nuclear collisions, such power-suppressed
collisions can be enhanced by the nuclear size, as in
Eq.~(\ref{param}).

\subsection{Factorization at Nonleading Powers}

Much of the predictive content of pQCD is contained in factorization
theorems like Eq.~(\ref{twist2conv}).  
In order to consistently treat the power
suppressed multiple scattering, we need a corresponding factorization
theorem for higher-twist (i.e., power suppressed) contributions to
hadronic hard scattering. 

\begin{figure}
\begin{minipage}[t]{3in}
\begin{center}
\epsfig{figure=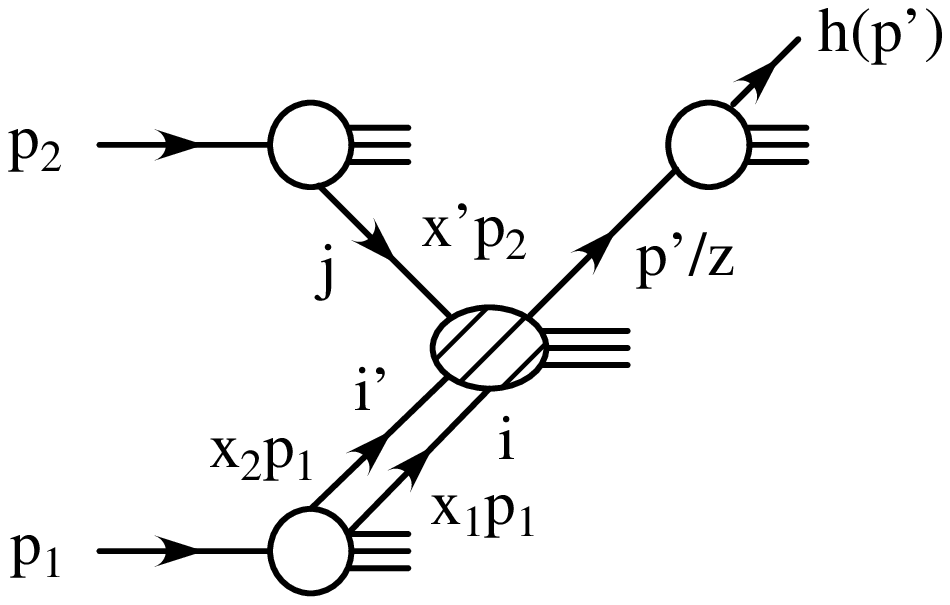,width=2.6in}
\end{center}
\caption{Scattering amplitude between two incoming hadrons with two
hard collision.}
\label{fig6}
\end{minipage}
\hfill
\begin{minipage}[t]{3in}
\begin{center}
\epsfig{figure=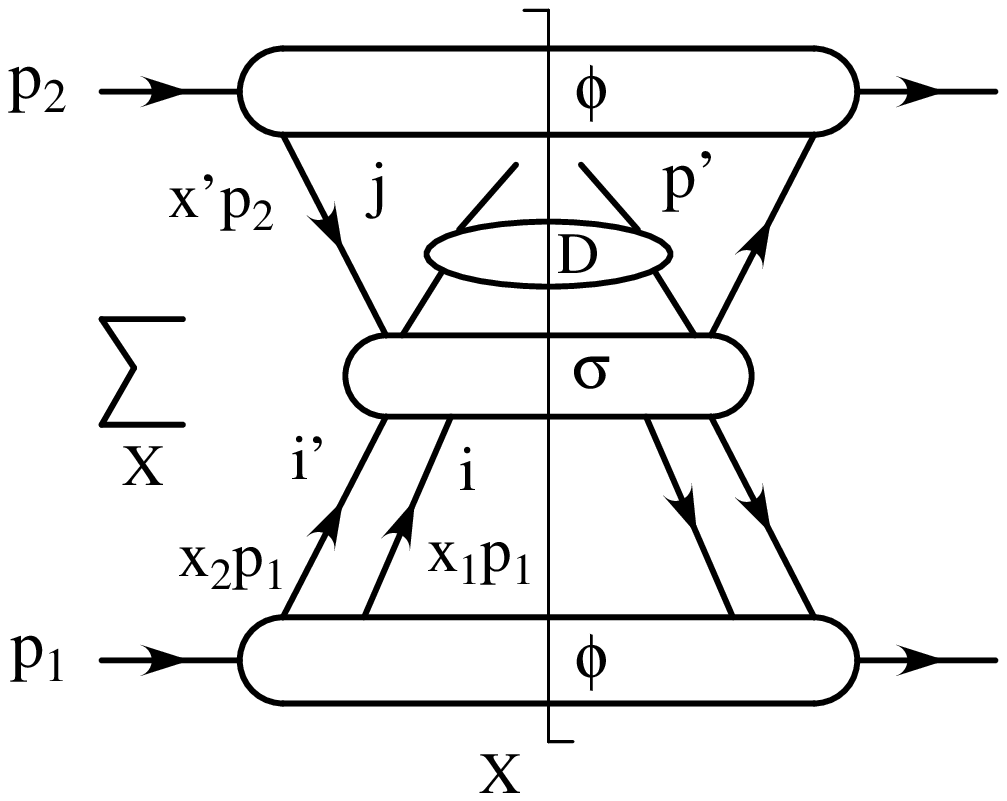,width=2.5in}
\end{center}
\caption{Perturbative QCD factorization at the next-to-leading power.}
\label{fig7}
\end{minipage}
\end{figure}

Fig.~\ref{fig6} is a picture for a power suppressed contribution
to hard scattering.  In this case, {\sl two} partons $i$ and $i'$
with momenta $x_1p_1$ and $x_2p_1$ from the target (the ``nucleus")
collide with a single parton $j$ of momentum $x'p_2$ (from the
``projectile").
After squaring the scattering amplitude in Fig.~\ref{fig6}, and
summing over all final states, the power suppressed contribution to
the cross section can be factorized into the form \cite{QS}
\beqa
\omega{d\sigma_4\over d^3p'}
&=&
\sum_{(ii')jk}\int\frac{dx'}{x'}\,f_{j/p_2}(x')\,
               \int\frac{dz}{z^2}\,D_{h/k}(z)\,
\noindent\\
&\ & \times
\int dx_1 dx_2 dx_3\; T_{(ii')/p_1}(x_1,x_2,x_3)\,
\hat{\sigma}^{(4)}_{(ii')+j}(x_ip_1,x'p_2,p'/z)\, ,
\label{sig4}
\eeqa
which can be illustrated by the sketch in Fig.~\ref{fig7}.
The expectation value $T$
corresponding to this multiparton contribution from the
target is typically of the form \cite{QS},
\beq
T_{(ii')/p}(x_1,x_2,x_3,Q)= \int{dy^-_1dy^-_2dy^-_3\over(2\pi)^3}
{\rm e}^{ip^+(x_1y^-_1+x_2y^-_2+x_3y^-_3)}
\langle p|B^\dagger_{i}(0)B^\dagger_{i'}(y^-_3)B_{i'}(y^-_2)B_{i}(y^-_1)
|p\rangle\, ,
\label{Tdef}
\eeq
where $B_i$ is the field corresponding to a parton of type
$i=q,{\bar q},G$.  In Eq.~(\ref{sig4}),
the hard part $\hat{\sigma}^{(4)}_{(ii')+j}$  depends on the
identities and momentum fractions of the incoming partons,
but is otherwise independent of the structure -- in particular
the size -- of the target (and projectile).
To find $A$-enhancement due
to multiple scattering, we must look elsewhere.

Before identifying the source of the $A$-enhancement, 
we briefly explain why the
factorization formula in Eq.~(\ref{sig4}) can be valid.
Although the formal arguments for the validity are well-documented
\cite{QS}, a heuristic understanding of factorization can be
useful.  Such a understanding may be found in the
Lorentz transformation properties of gauge fields \cite{BRS-twist4}.

\begin{figure}
\begin{minipage}[t]{2.4in}
\begin{center}
\epsfig{figure=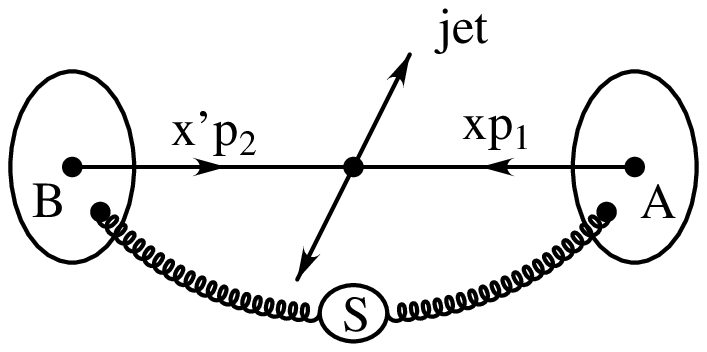,width=2.2in}
\end{center}
\caption{Interactions of long-range fields that might break the
factorization.}
\label{fig8}
\end{minipage}
\hfill
\begin{minipage}[t]{3.8in}
\begin{center}
\epsfig{figure=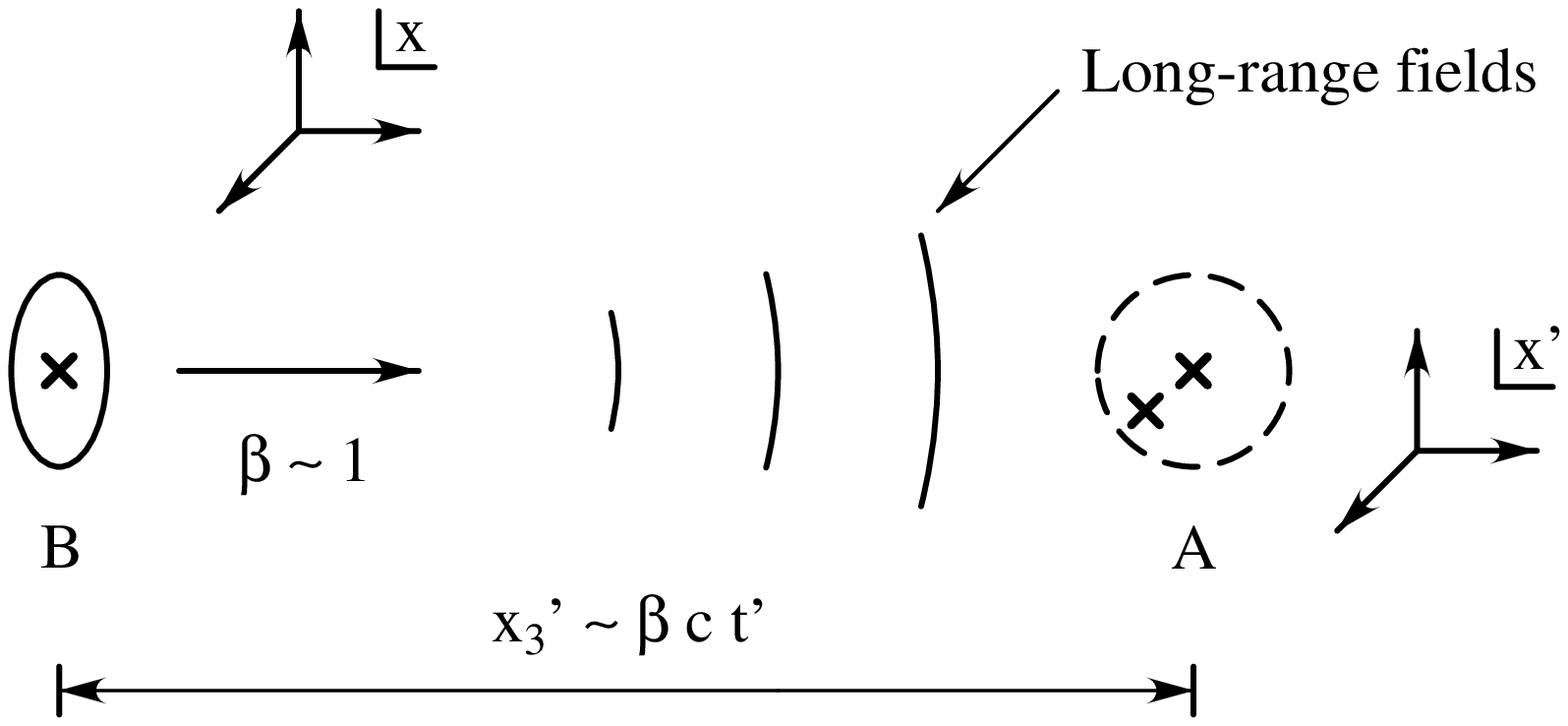,width=3.8in}
\end{center}
\caption{Sketch for soft interactions of long-range fields between two
colliding hadrons $A$ and $B$.}
\label{fig9}
\end{minipage}
\end{figure}

In hadron-hadron collisions, the factorization could be broken if
interactions of long-range fields, labeled by ``$S$'' in
Fig.~\ref{fig8}, between the two incoming hadrons are 
important.  The interactions of
the long-range fields could alter the hadronic states of the incoming
hadrons, and subsequently, change the parton matrix elements (or
distributions) before the hard collisions take place.
Without universality of the parton matrix elements (or
distributions) in Eqs.~(\ref{twist2conv}) and (\ref{sig4}), the
factorized formulas would lose predictive power, and we would say that 
factorization is broken.  
Formal proofs of the factorization theorem 
must show that all such long-range soft interactions are
either power suppressed, or can be effectively removed due to
unitarity, for inclusive observables \cite{CSS-fac}.

Let us now review a heuristic argument 
that soft interactions between the two incoming hadrons are
kinematically suppressed, due to the Lorentz transformation properties
of gauge fields \cite{BRS-twist4}.  Consider a collision between
hadron $A$ and hadron $B$, as shown in Fig.~\ref{fig9}.  Hadron
$B$ is moving very fast along the direction of $x_3$ with $\beta\sim
1$, while hadron $A$ is at rest in the $x'$ frame.
Let us define $\Delta\equiv \beta c t' - x_3'$, 
where $x_3'$ is the third component in the $x'$ frame.  
The long-range fields generated by hadron $B$, at rest in the $x$
frame, behave very differently in the $x$ and $x'$ frames, and
different types of fields have different properties 
under Lorentz transformations.  For example, the three
types of fields: scalar field, longitudinal component of a gauge
field, and its $E_3$ field strength, have the following behaviors;
with $\gamma = (1-\beta^2)^{-1/2}$,
\beqa
\mbox{\underline{Field}} \quad \
  & \quad \ \mbox{\underline{$x$-Frame}} \quad \
  & \quad \
    \mbox{\underline{$x'$-Frame}}
\nonumber \\
  &
  &
\nonumber \\
\mbox{Scalar} \quad \
  & \quad \
    V(x) = \frac{e}
                {\left|\vec{x}\right|}
    \quad \
  & \quad \
    V'(x') = \frac{e}
                  {(x_T^2+\gamma^2\Delta^2)^{1/2}}
\nonumber \\
  &
  & \quad \
    \Longrightarrow \frac{1}{\gamma} \quad
                    \mbox{``contracted like a ruler''}
\label{scalar} \\
\mbox{Longitudinal Gauge} \quad \
  & \quad \
    A^-(x) = \frac{e}{|\vec{x}|}
    \quad \
  & \quad \
    A'^-(x') = \frac{e\gamma(1+\beta)}
                    {(x_T^2+\gamma^2\Delta^2)^{1/2}}
    \quad \
\nonumber \\
  &
  & \quad \
    \Longrightarrow 1 \quad
                    \mbox{``not contracted!''}
\label{gauge} \\
\mbox{Field Strength}
    \quad \
  & \quad \
    E_3(x) = \frac{e}{|\vec{x}|^2}
    \quad \
  & \quad \
    E_3(x') = \frac{-e\gamma\Delta}
                    {(x_T^2+\gamma^2\Delta^2)^{3/2}}
\nonumber \\
  &
  & \quad \
    \Longrightarrow \frac{1}{\gamma^2} \quad
                    \mbox{``strongly contracted!''}
\label{field}
\eeqa
Although the magnitude of the longitudinal component $A'^-(x')$ 
of the gauge field is not suppressed under the
Lorentz transformation, as shown in Eq.~(\ref{gauge}), and its
interactions can be very strong, a short calculation shows that as
$\beta\rightarrow 1$, it becomes gauge equivalent to a vanishing gauge
field.  On the other hand, the gluon field strength is 
suppressed under the Lorentz transformation even more strongly than
the scalar field, Eq.~(\ref{scalar}).  
In terms of energy scales, the $1/\gamma^2$ in
Eq.~(\ref{field}) translates into
a suppression factor of $1/Q^4$, which suggests that the factorization
should be valid at the order of $1/Q^2$, and might fail at $O(1/Q^4)$
\cite{QS,BRS-twist4,DFT-Q4,BFT-Q4}.

Showing the factorization at the next-to-leading power is a beginning
toward a unified discussion of $O(1/Q^2)$ effects in a wide class of
processes.  A systematic treatment of double scattering in a nuclear
medium is an immediate application of the generalized factorization
theorem.

\subsection{$A$-Enhancement from Matrix Elements}

As we pointed out in last subsection, the partonic hard part in the
factorized formula in Eq.~(\ref{sig4}) is independent of the
structure -- in particular the size -- of the target.  Therefore, we
need to find the $A$-enhancement due to multiple scattering from the
matrix elements, if there is to be any size enhancement.

For definiteness, we  consider photoproduction or deeply inelastic
scattering on a nucleus \cite{LQS1,LQS3}.  In this case, the
additional soft scattering is
always a final-state interaction.
The structure of the target is manifest only in
the matrix element $T$ in Eq.~(\ref{sig4}).
Each pair of fields in the matrix element Eq.~(\ref{Tdef}) represents
a parton that participates in the hard scattering.
The $y^-_i$ integrals parameterize the distance between
the positions of these particles along the path of the
outgoing scattered quark.
In Eq.~(\ref{Tdef}), integrals over the
distances $y^-_i$ generally cannot grow with the size
of the target because of oscillations from the exponential factors
${\rm e}^{ip^+x_iy^-_i}$.

Since the kinematics of a single-scale hard collision is only
sensitive to the total momentum from the target, two of the three
momentum fractions: $x_1,x_2$, and $x_3$ cannot be fixed by the
hard collisions.  If the integration of the momentum fractions is
dominated by the region where $x_2\sim 0$ and $x_3\sim 0$, the
corresponding $y_i^-$ integration in Eq.~(\ref{Tdef}),
$$
\left.
\int dy_i^-\ {\rm e}^{ix_i p^+ y_i^-}\right|_{x_i\sim 0}
\propto\ \mbox{size of the target},
$$
is proportional to the size of the target or the $A^{1/3}$.

\begin{figure}
\begin{center}
\epsfig{figure=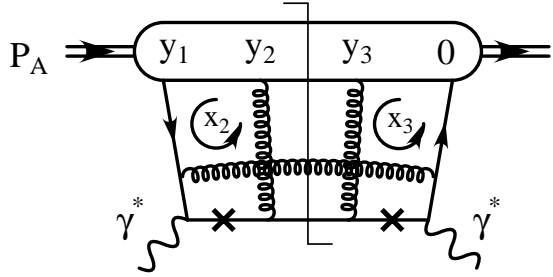,width=2.9in}
\end{center}
\caption{Poles that give rise to an $A$-enhanced cross section in
deeply inelastic scattering.}
\label{fig10}
\end{figure}

The question now is if it is possible that the partonic parts,
$\hat{\sigma}_{(ii')+j}^{(4)}$ in Eq.~(\ref{sig4}) can have
contributions that are dominated by regions where one or two of the
three momentum fractions vanish.  The answer is yes.  Some of the
$\hat{\sigma}_{(ii')+j}^{(4)}$ in the $x_i$ integrals have two
poles, labeled by the crosses ``$\times$'' in
Fig.~\ref{fig10}.
The symbols ``$\times$'' in Fig.~\ref{fig10} represent the potential
poles when the corresponding propagators become on-shell.  In the
example of deeply inelastic scattering, the poles are always
associated with the scattered particle \cite{LQS1,LQS3}, while in
other processes, the poles can also be associated with incoming
particles \cite{GQ,Guo:1998rd}.
It is important to emphasize that using
a pole in the complex $x_i$ (longitudinal momentum)
space to do the integral does not correspond to
assuming on-shell propagation for the
scattered quark.  Indeed, the $x_i$ integrals are not pinched
between coalescing singularities at such points, and the same
results could be derived by performing the $x_i$ integrals
without ever going through the $x_i=0$ points.
It is also worth noting that this is a feature unique to higher-twist
matrix elements, for which the $x_i$ are not restricted to be positive
definite.  Physically, this is possible because, unlike leading twist
matrix elements, they do not generally have the interpretations of
probabilities. 

\begin{figure}
\begin{minipage}[t]{3in}
\begin{center}
\epsfig{figure=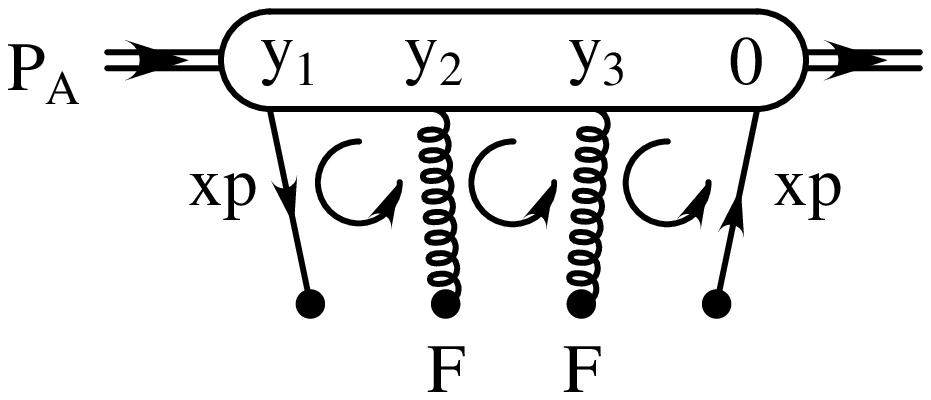,width=2.6in}
\end{center}
\caption{Sketch of the twist-4 quark-gluon correlation function that 
gives rise to an $A^{1/3}$ type enhancement.}
\label{fig11}
\end{minipage}
\hfill
\begin{minipage}[t]{3in}
\begin{center}
\epsfig{figure=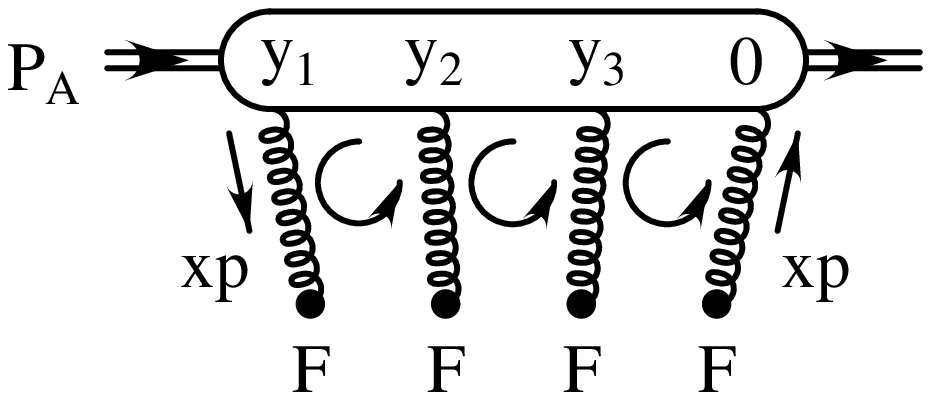,width=2.6in}
\end{center}
\caption{Sketch of the twist-4 gluon-gluon correlation function that
gives rise to an $A^{1/3}$ type enhancement.}
\label{fig12}
\end{minipage}
\end{figure}

The result of this reasoning is that two of the three momentum
fraction integrations: $dx_1\,dx_2\,dx_3$ in Eq.~(\ref{sig4}) are
fixed by the two poles, and the convolution over
$dx_1\,dx_2\,dx_3$ in Eq.~(\ref{sig4}) is
simplified to an integration over only one momentum fraction,
\beqa
& \ &
\int dx_1\,dx_2\,dx_3\, T_{(ii')/p_1}(x_1,x_2,x_3)\,
        \hat{\sigma}^{(4)}_{(ii')+j}(x_ip_1,x'p_2,p'/z)
\nonumber \\
& \ & \Longrightarrow
\int dx\, T_{q}(x,A)\, \hat{\sigma}^{(D)}_{(q)+j}(xp_1,x'p_2,p'/z),
\label{poles2x}
\eeqa
where the partonic part $\hat{\sigma}^{(D)}$ is finite and
perturbative with the superscript $(D)$ indicating the contribution
from double scattering.  The above matrix element, $T_{q}(x,A)$, as
illustrated in Fig.~\ref{fig11},
has the form
\begin{eqnarray}
T_{q}(x,A) &=& \int {\d y^-_1 \over 2\pi}
e^{ip^+xy_1^-}
\int {\d y^-_2 \d y^-_3 \over 2\pi}
\theta(y^-_2-y_1^-)\theta(y_3^-) \nonumber
\\
&\ &\quad \times\; {1 \over 2}\; \langle p_A|{\bar q}(0)
\gamma^+ F^{\alpha +}(y_3^-)
{F^{+}}_\alpha(y^-_2) q(y_1^-)|p_A\rangle\ ,
\label{qkme}
\end{eqnarray}
where $|p_A\rangle$ is the relevant nuclear state.
The variable $x$ here is the fractional momentum associated with the
hard parton from the target that initiates the process.
The soft scattering contributes a negligible longitudinal fractional momentum.
Details of the reasoning and calculation for deeply inelastic scattering
are given in Ref.~\cite{LQS3}.

Similar to the quark-gluon correlation function $T_{q}(x,A)$ in
Eq.~(\ref{qkme}), another important twist-4 parton correlation
function that gives rise to an $A^{1/3}$ enhancement is the
gluon-gluon correlation function \cite{LQS1},
\begin{eqnarray}
T_{g}(x,A) &=& \int {\d y^-_1 \over 2\pi}
e^{ip^+xy_1^-}
\int {\d y^-_2 \d y^-_3 \over 2\pi}
\theta(y^-_2-y_1^-)\theta(y_3^-)
\nonumber \\
&\ &\quad \times\; {1 \over xp^+}\; \langle p_A|F^{\beta+}(0)
F^{\alpha +}(y_3^-) {F^{+}}_\alpha(y^-_2) {F^{+}}_\beta(y_1^-)
|p_A\rangle\ ,
\label{gkme}
\end{eqnarray}
which is illustrated in Fig.~\ref{fig12}.

In this form of the twist-4 parton-parton correlation functions, two
integrals over the $y^-$ and $y_2^-$ can grow with the nuclear radius
as fast as $A^{1/3}$.
However, if we require local color confinement, the
difference between the light-cone coordinates of the two field
strengths should be limited to the nucleon size.  Therefore, only one
of the two $y_i^-$ integrals can be extended to the size of nuclear
target.  The twist-4 parton-parton correlation functions are then
proportional to the size of the target, that is, 
enhanced by $A^{1/3}$.

\section{Applications}

In Refs.\ \cite{LQS1} and
\cite{LQS3}, we have applied the formalism sketched above to single-particle
inclusive and single-jet production for
deeply inelastic scattering and photoproduction.  These
cases involve final-state interactions only.
In each case, the leading $1/Q^2$ correction is proportional to the
matrix elements  in Eqs.~(\ref{qkme}) and (\ref{gkme}).
Of course, the value of the correction cannot be estimated
without an idea of the magnitudes of the $T$'s.  Since these magnitudes
are nonperturbative they must be taken from experiment.
At the same time, we expect the $x$-dependence of
the probability to detect  the hard parton to
be essentially unaffected by the presence or  absence of an
additional soft scattering.  Thus, we choose the ansatz \cite{LQS1}
\beq
T_i(x,A)=\lambda^2\, A^{1/3}\, f_{i/p_A}(x,A)
\label{Tqans}
\eeq
for $i=q,g$ in terms of the corresponding twist-two effective nuclear
parton distribution $f_{i/A}$, with $\lambda$ a constant with
dimensions of mass (see Eq.\ ({\ref{param})).
This assumption facilitates the comparison to data.

\subsection{Momentum Imbalance of Di-Jets in Photoproduction}

A quantity that is sensitive to final-state rescattering in a
particularly direct way is the momentum imbalance of di-jets in
photoproduction in nuclei.  The $A^{4/3}$ dependence of this
quantity measured by the Fermilab E683 collaboration is a clear signal
of the presence of double scattering \cite{E683}.

The di-jet momentum imbalance $k_{T\phi}$ measured at Fermilab is
defined as \cite{E683}
\beq
k_{T\phi} \equiv \frac{1}{2} \left(p_{T_1}+p_{T_2}\right)
                  \sin \Delta\phi\, ,
\label{ktphi}
\eeq
where $p_{T_1}$ and $p_{T_2}$ are the transverse momenta of the
di-jets and the angle between the di-jets $\Delta\phi$ is defined in 
Fig.~\ref{fig13}.
\begin{figure}
\begin{minipage}[t]{2.9in}
\begin{center}
\epsfig{figure=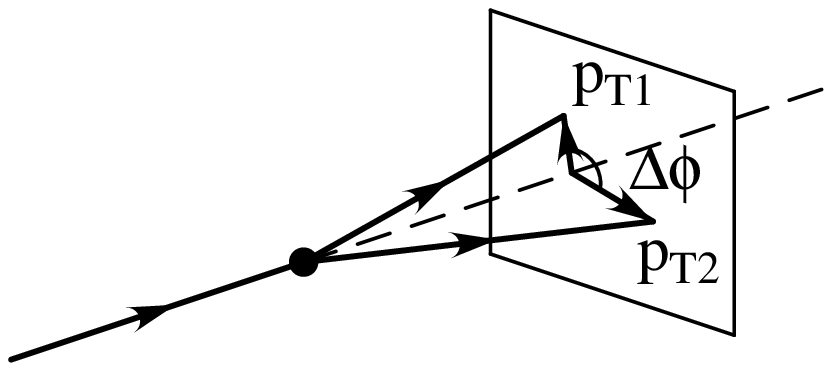,width=2.8in}
\end{center}
\caption{Sketch for the di-jet momentum imbalance measured at
Fermilab E683.}
\label{fig13}
\end{minipage}
\hfill
\begin{minipage}[t]{3.2in}
\begin{center}
\epsfig{figure=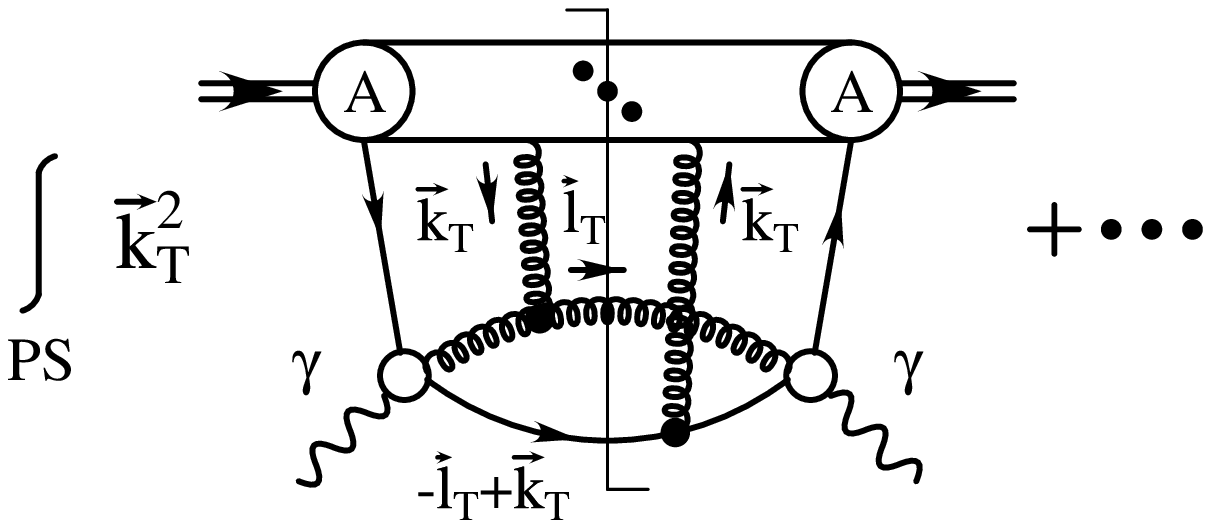,width=3.2in}
\end{center}
\caption{Sketch for the lowest order contributions to the averaged
di-jet momentum imbalance in photoproduction.}
\label{fig14}
\end{minipage}
\end{figure}
This momentum imbalance is the
projection of the momentum imbalance out of the scattering plane
defined by the beam axes and either of the two observed jets.
The data from E683 shows that the averaged momentum imbalance $\langle
k_{T\phi}^2 \rangle$ can be parameterized as
$(1.44+0.174A^{1/3})$~GeV$^2$.  The term proportional to $A^{1/3}$
should be a consequence of the multiple scattering of these two jets
when they pass through the nuclear medium \cite{LQS}.

In Ref.~\cite{LQS}, we calculated the di-jet momentum imbalance in
pQCD by applying the twist-4 factorization formalism to double
scattering in a nuclear medium.  The lowest-order contributions to the
momentum imbalance are shown schematically in Fig.~\ref{fig14}.
At the lowest order, the di-jet momentum imbalance is approximately
equal to the momentum imbalance between the two scattered partons
which fragment into the two leading jets: $k_T^2\approx \hat{k}_T^2$.
As shown in Figs.~\ref{fig13} and 14, the averaged momentum imbalance
$\langle \hat{k}_T^2 \rangle$ is not exactly the same as the
$\langle k_{T\phi}^2\rangle$ used in the experiment.  The relation,
assuming rotational symmetry, is simply (we always use rms averages)
\beq
\langle k_T^2\rangle = 2 \langle k_{T\phi}^2 \rangle\, .
\label{kt2-ktphi2}
\eeq
Letting $d\hat{\sigma}^{\gamma i}(xp,\ell,p_\gamma)$ 
denote the photon-parton Born
cross section with parton flavor ($i=q,\bar{q},g$) for fixed leading
jet momentum $\ell$, we derived the double scattering contribution to
the momentum imbalance \cite{LQS},
\beqa
\langle k_T^2 E_\ell \frac{\d\sigma^{\gamma A}(\ell)}{d^3\ell}
\rangle_{4/3}
&=&
\lambda^2\, A^{4/3}\, \pi^2\alpha_s \left[
C_F\sum_{i=q.\bar{q}} \int dx f_{i/A}(x)
        E_\ell \frac{\d\hat{\sigma}^{\gamma i}}{d^3\ell}
        (xp,\ell,p_\gamma)
\right.
\nonumber\\
&\ & {\hskip 0.8in} \left.
+ C_G \int dx f_{g/A}(x)
        E_\ell \frac{\d\hat{\sigma}^{\gamma g}}{d^3\ell}
        (xp,\ell,p_\gamma)
\right]
\label{dijets}
\eeqa
with $C_F=4/3$ and $C_G=3$.  The average value of $k_T^2$ for all
events in a region $R$ of dijet phase space is found from
Eq.~(\ref{dijets}) by integrating over that region and dividing by the
corresponding un-weighted cross section \cite{LQS}.  If we assume that
$\alpha_s(\ell_T^2)$ in Eq.~(\ref{dijets}) is evaluated at a typical
value of the momentum transfer in $R$, and is kept approximately
constant within $R$, the averaged momentum imbalance at the lowest
order can be expressed as
\beq
\langle k_T^2(R)\rangle_{4/3}
= \lambda^2\, A^{4/3}\, \pi^2\alpha_s \,
\frac{C_F\sigma_q^{\gamma A}(R,p_\gamma)
      +C_G\sigma_g^{\gamma A}(R,p_\gamma)}
      {\sigma_q^{\gamma A}(R,p_\gamma)
      +\sigma_g^{\gamma A}(R,p_\gamma)}\, .
\label{kt2-R}
\eeq
In this expression, the un-weighted cross sections are defined as
\beq
\sigma_i^{\gamma A}(R,p_\gamma)
=\int_R\, \frac{d^3\ell}{E_\ell(2\pi)^3}\,
         \int dx\, f_{i/A}(x)\,
         E_\ell \frac{d\hat{\sigma}^{\gamma i}}{d^3\ell}
         (xp,\ell,p_\gamma)\, .
\label{cross-x-R}
\eeq
By comparing Eq.\ (\ref{kt2-R}) to data \cite{E683}, we
found $\lambda^2\sim 0.05-0.1$ GeV$^2$ \cite{LQS}.  This value
may be used to predict anomalous $A$-enhancement for other
processes.

The lowest-order calculation given above neglects the evolution of the
final state partons into jets.  
We have argued above, however, that the evolution of the jet
takes place outside of the nucleus.   It is interesting to see how this
decoupling of jet evolution occurs in our calculation.

In perturbation theory, higher-order corrections due to the evolution of the
final state partons in
Fig.~\ref{fig14} arise from diagrams of the 
kind shown in Fig.~\ref{fig15}.
\begin{figure}
\begin{minipage}[t]{2.6in}
\begin{center}
\epsfig{figure=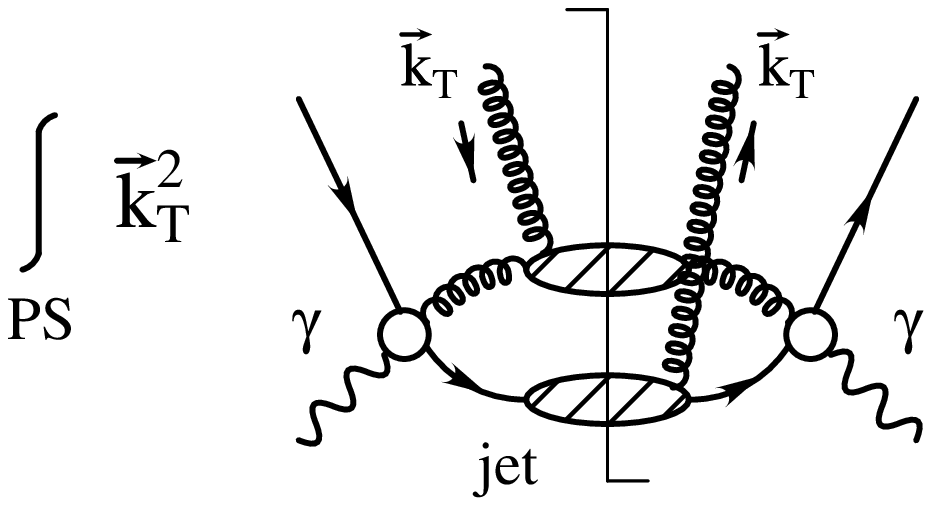,width=2.4in}
\end{center}
\caption{Sketch for the high order contributions to di-jets momentum
imbalance in photoproduction.}
\label{fig15}
\end{minipage}
\hfill
\begin{minipage}[t]{3.5in}
\begin{center}
\epsfig{figure=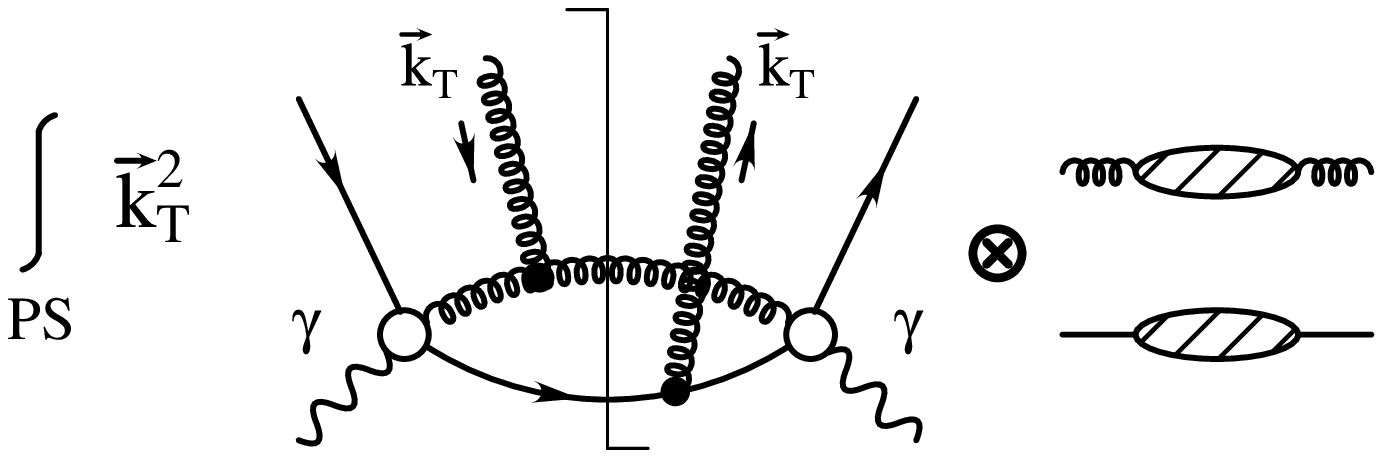,width=3.4in}
\end{center}
\caption{Sketch for the factorization of high order parton evolutions
to jets from the lowest order parton level double scattering.}
\label{fig16}
\end{minipage}
\end{figure}
The shaded subdiagrams may be thought of as
``jet functions", which represents the evolution of the outgoing partons. 
The first thing to note is that
leading corrections due to final-state partonic 
bremmstrahlung are from small angle emission, which stays within the
respective jets.  Such collinear emission does not change 
the imbalance, $k_T$, although 
any of these fragmentation products
can, in principle, exchange soft gluons with the nucleus.
As pointed out in Ref.~\cite{LQS}, however,
soft gluon multiple scattering cannot resolve the details of the jets.
The coupling of the final-state jet to soft gluons is coherent, and
equivalent to a single ``eikonal'' line at leading power,
whose direction and color are defined by the parton 
that initiates the jet.
Thus, soft nuclear rescatterings factor from the jets, as
shown in Fig.~\ref{fig16}.  The combination of two
such eikonal couplings, weighted by $k_T^2$ is enough
to generate a lowest-order contribution to the matrix
elements of Eqs.\ (\ref{qkme}) or (\ref{gkme}).
At yet higher orders, but
remaining at first nonleading power in $Q$, additional
soft gluon corrections generate nonabelian phases  \cite{CSS-fac},
(ordered exponentials of the gauge field), which 
serve to make the higher-twist matrix elements
gauge-invariant, but do not otherwise affect $A$-dependence.
One way of looking at this is that the di-jet $k_T$ imbalance
is relatively insensitive to energy loss.  Thus, the
leading dependence of Eq.\ (\ref{param}) is stable all orders in 
perturbation theory.

\subsection{$A$-Enhancement in Other Processes}

In order to test the theory, we need to identify other physical
observables that are sensitive to the same twist-4 parton-parton
correlation functions $T_i(x,A)$ with $i=q,g$.
One such process is direct photon production at measured transverse
momentum, whose very moderate $A$-dependence has been measured by the
E706 experiment at Fermilab.  In Ref.~\cite{GQ}, it was
found that the value of $\lambda^2$ above, which produces a relatively
large enhancement in
dijet momentum imbalance, due to final-state interactions,
produces a quite small $A$-enhancement
in photoproduction, due to initial-state interactions,
consistent with experiment.
This may shed some light on the long-standing observation that
(initial-state) transverse
momentum effects in Drell-Yan cross sections, to which we now turn,
are also surprisingly small \cite{DY,E772-dy,NA10-dy}.

In Ref.~\cite{Guo:1998rd}, Drell-Yan transverse momentum broadening
was calculated at the lowest order in pQCD.  By evaluating the
lowest order diagram in Fig.~\ref{fig17} plus corresponding
interference diagrams, it was found that the Drell-Yan transverse
momentum broadening in hadron-nucleus collisions can be
expressed in terms of the same twist-4 quark-gluon correlation
function $T_q(x,A)$ \cite{Guo:1998rd},
\beq
\langle Q_T^2 \rangle_{4/3}
= \left(\frac{4\pi^2 \alpha_s}{3}\right)\,
\frac{\sum_q e_q^2 \int dx' f_{\bar{q}/h}(x')\, T_q(\tau/x',A)/x'}
      {\sum_q e_q^2 \int dx' f_{\bar{q}/h}(x')\, f_q(\tau/x',A)/x'}
\label{dy-qt2}
\eeq
where $\sum_q$ runs over all quark and antiquark flavors, $e_q$ is
the quark fractional charge, and $\tau=Q^2/s$, in terms of the
lepton-pair invariant mass $Q$ and hadron-hadron center of mass
energy, $\sqrt{s}$. 
\begin{figure}
\begin{minipage}[t]{2.8in}
\begin{center}
\epsfig{figure=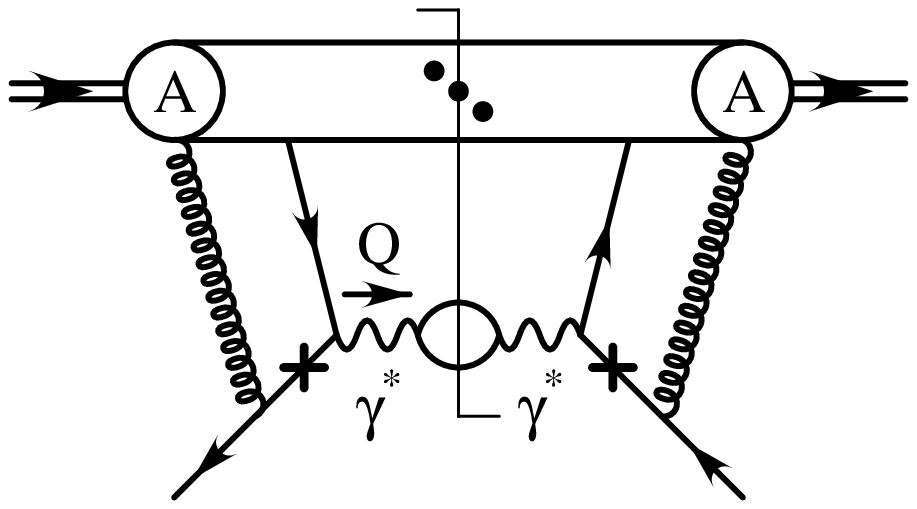,width=2.5in}
\end{center}
\caption{Lowest order diagram that contributes to the Drell-Yan
transverse momentum broadening in hadron-nucleus collisions.}
\label{fig17}
\end{minipage}
\hfill
\begin{minipage}[t]{3.3in}
\begin{center}
\epsfig{figure=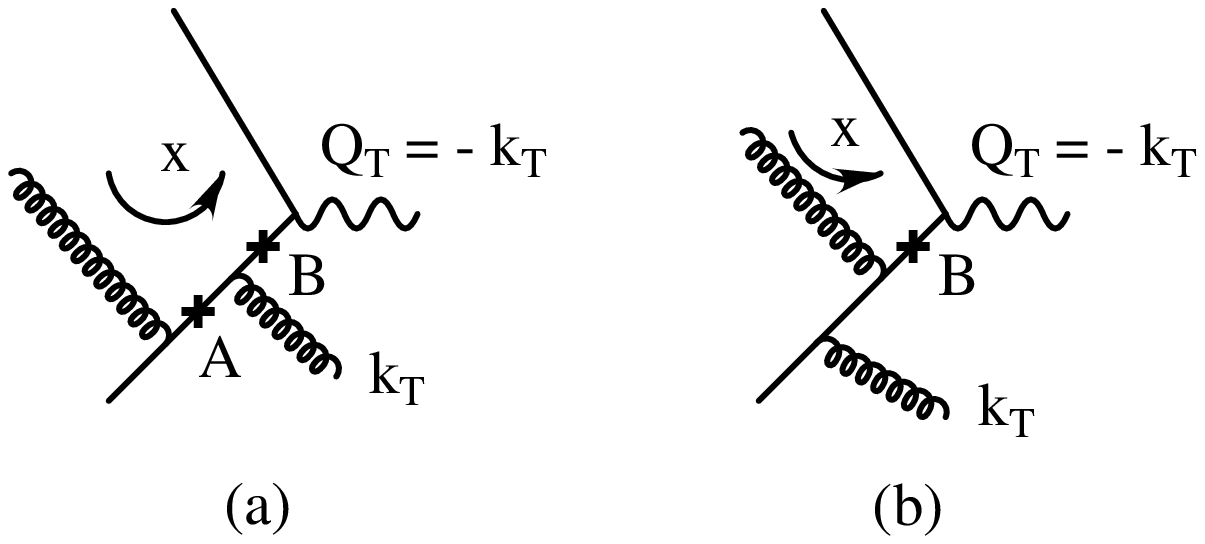,width=3.3in}
\end{center}
\caption{Sample double scattering amplitudes that contribute to the
NLO $A$-enhancement.}
\label{fig18}
\end{minipage}
\end{figure}
Adopting the model in Eq.~(\ref{Tqans}), the lowest order Drell-Yan
transverse momentum broadening in Eq.~(\ref{dy-qt2}) can be simplified
as
\beq
\langle Q_T^2 \rangle_{4/3}
= \left(\frac{4\pi^2 \alpha_s}{3}\right)\,
   \lambda^2\,A^{1/3}\, .
\label{dy-qt2-lqs}
\eeq
By comparing Eq.~(\ref{dy-qt2-lqs}) to data from Fermilab E772 and
CERN NA10 experiments \cite{E772-dy,NA10-dy}, it was found
\cite{Guo:1998rd} that Drell-Yan data favor a $\lambda$ value:
$\lambda_{\rm DY}^2 \sim 0.01$~GeV$^2$, which is considerably smaller
than the $\lambda^2$ extracted from the dijet momentum imbalance in
photoproduction.

It could be that the source of this discrepancy is experimental.  On
the other hand, we should also consider possible differences between
the roles of multiple scattering in dijet momentum imbalance and
Drell-Yan.  For dijet momentum imbalance in photoproduction, soft
gluon multiple scatterings are purely final-state interactions, while
the Drell-Yan transverse momentum broadening at the lowest order are
caused by purely initial-state interactions.  For initial-state
interactions, there is strong interference beyond the leading order.
Consider the diagrams in Fig.~\ref{fig18}, which contribute to the
Drell-Yan transverse momentum broadening,
\beq
\langle Q_T^2 \rangle_A
= A\, \langle Q_T^2 \rangle_{A=1}
+ \langle Q_T^2 \rangle_{4/3}
\label{dy-qt2-tot}
\eeq
at the next-to-leading order.  Actually, these two diagrams can
contribute to both terms in Eq.~(\ref{dy-qt2-tot}).  In order to
extract the nuclear size-enhanced contribution to the broadening 
(the $A^{4/3}$ term), we study the behavior of these 
diagrams at the poles near $x=0$.  
For the diagram in Fig.~\ref{fig18}a,
there are two potential poles labeled by $A$ and $B$.  
When $k_T$ is small, contributions from these two poles cancel. 
That is, the NLO contributions to 
the Drell-Yan transverse momentum spectrum from this
diagram do not have the dominant $1/Q_T^2$ behavior, 
and therefore, its
contributions to the $\langle Q_T^2 \rangle_{4/3}$ is much suppressed.
When $k_T$ is small, the diagram in Fig.~\ref{fig18}b is already
included in the lowest order contribution to the broadening, because
the gluon radiation is a part of the parton distribution
$f_{\bar{q}/h}$.  On the other hand, when $k_T$ is large, the only
possible pole at $B$ requires a finite value of $x$, which ruins the
$A$-enhancement \cite{Guo:1998rd}.

From the above example, it seems plausible 
to us that the interference of
initial-state radiation and multiple scattering suppresses $A$-enhancement
in Drell-Yan cross sections.  
By contrast, 
the corresponding effects of final-state radiation may cancel
for an inclusive jet cross section, because we do not observe the $k_T$
of radiation within the high-$p_T$ jets.  For Drell-Yan,
the two terms of Eq.\ (\ref{dy-qt2-tot}) are both from initial-state
interactions, while in the dijet case,
the first is primarily from the initial state, while the
second is final-state.   In the Drell-Yan case, destructive interference
between the two mechanisms is possible, for di-jets, it is not.
On this basis, we may expect
a stronger $A$-enhancement to the dijet momentum imbalance in
photoproduction than to the transverse momentum in Drell-Yan.  
Clearly, further study of this difference between initial-state and
final-state interactions and related questions is in order.

\section{Conclusions}

In conclusion, we have argued that the nuclear size (or
$A^{1/3}$-type) enhancement caused by multiple scattering can be
consistently calculated in pQCD, in terms of generalized
factorization theorems. By studying enhancement from the nuclear medium,
we can learn about strong interaction physics at twist-4.  
The first power correction measures {\it new} matrix
elements $\langle q(FF)q\rangle$ and $\langle F(FF)F\rangle$.
These matrix elements
provide new insights into the nonperturbative regime of QCD.

As reviewed above, the initial applications of pQCD to the first power
corrections are interesting and generally successful.  Different observables,
such as jet broadening in deeply inelastic scattering
\cite{Guo:1998rd} and pion transverse momentum broadening in deeply
inelastic scattering \cite{Guo:2000eu}, have been proposed and
evaluated.  The same techniques have been recently applied to many new
physical observables \cite{GW-loss,Guo-DY-pt,Fries:1999jj,Huang:1999cm,
Fries:2000da,GQZ-DY,GZZ-DY-xf,SL-gamma}.

\section*{Acknowledgments}

The work of G.S.\ was supported in part by the National Science Foundation,
grants PHY9722101 and PHY0098527.
The research of J.Q.\ at Iowa State was supported in part by
the US Department of Energy under Grant No. DE-FG02-87ER40371.


\end{document}